\documentclass[twocolumn,twocolappendix]{aastex7}
\usepackage{mhchem}
\usepackage{leftindex}
\usepackage[colorinlistoftodos]{todonotes}
\usepackage{amssymb}
\usepackage{comment}
\usepackage{multirow,acronym}
\usepackage{natbib}
\usepackage{listings} 
\usepackage{makecell}
\usepackage{booktabs}
\usepackage{amsmath}
\usepackage{subfigure}
\usepackage{color}
\usepackage{xcolor}
\usepackage{hyperref}
\usepackage[T1]{fontenc}
\usepackage{graphicx} 
\usepackage{bm}
\hypersetup{colorlinks=true, citecolor=blue, 
  linkcolor=cyan, urlcolor=magenta} 
\usepackage{aas_macros}
\usepackage{amsfonts}
\usepackage{float}
\usepackage{amsmath,amssymb}
\usepackage{natbib}    % For use with bibtex
\usepackage{hyperref} % For internal hyperlinks
\usepackage{graphicx} % For included graphics
\usepackage{color}
\usepackage{verbatim}
\usepackage{enumitem}
\usepackage{natbib}

\usepackage[linesnumbered,ruled]{algorithm2e}

\newcommand{\proptosim}{\mathrel{\vcenter{
 \offinterlineskip\halign{\hfil$##$\cr
 \propto\cr\noalign{\kern2pt}\sim\cr\noalign{\kern-2pt}}}}}
\renewcommand{\b}[1]{\boldsymbol{#1}}
\newcommand{\unit}[1]{{\rm #1}}

\newcommand{\response}[1]{{\color{blue} #1}}

\renewcommand{\max}{\mathrm{max}}
\newcommand{\au}{\mathrm{AU}}
\newcommand{\cm}{\unit{cm}}
\newcommand{\m}{\unit{m}}
\newcommand{\g}{\unit{g}}
\newcommand{\G}{\unit{G}}
\newcommand{\K}{\unit{K}}

\renewcommand{\micron}{\unit{\mu m}}

\newcommand{\eV}{\unit{eV}}

\newcommand{\s}{\mathrm{s}}

\newcommand{\ang}{\text{\AA}}

\newcommand{\B}{\mathbf{B}}     % Magnetic fields
\newcommand{\E}{\mathbf{E}}     % Electric fields
\newcommand{\J}{\mathbf{J}}     % Electric current fields
\renewcommand{\v}{\mathbf{v}}   % Velocity fields
\newcommand{\kb}{k_\mathrm{B}}

\renewcommand{\d}{\mathrm{d}}

\newcommand{\e}{\mathrm{e}}

\newcommand{\p}{\mathrm{p}}     % Poloidal
\newcommand{\ad}{\mathrm{ad}}   % adiabatic
\renewcommand{\O}{\mathrm{O}}     % Ohmic
\newcommand{\A}{\mathrm{A}}     % Ambipolar
\renewcommand{\H}{\mathrm{H}}     % Hall
\renewcommand{\P}{\mathrm{P}}     % P
\newcommand*\chem[1]{\ensuremath{\mathrm{#1}}}
\newcommand{\figdir}{.}

\newcommand{\DOA} {Department of Astronomy, School of
  Physics, Peking University, Beijing 100871, China}
\newcommand{\KIAA}{Kavli Institute for Astronomy and
  Astrophysics, Peking University, Beijing 100871, China}

\renewcommand{\response}[1]{#1}

\begin{document}

\title{Adsorption of volatiles on dust grains in
  protoplanetary disks}

\author[0000-0002-6540-7042]{Lile Wang}
\affil{\KIAA}
\affil{\DOA}
\email{lilew@pku.edu.cn}

\author[0000-0002-7607-719X]{Feng Long}
\affiliation{\KIAA}
\affiliation{\DOA}
\email{long.feng@pku.edu.cn}

\author[0000-0002-8537-6669]{Haifeng Yang}
\affiliation{Institute for Astronomy, School of Physics,
  Zhejiang University, 886 Yuhangtang Road, Hangzhou 310027,
  China}
\affiliation{Center for Cosmology and Computational
  Astrophysics, Institute for Advanced Study in Physics,
  \\ Zhejiang University, Hangzhou 310027, China}
\email{hfyang@zju.edu.cn}

\author[0000-0001-9290-7846]{Ruobing Dong}
\affiliation{\KIAA}
\email{rbdong@pku.edu.cn}

\author[0000-0001-7268-9917]{Shenzhen Xu}
\affil{School of Material Sciences,
  Peking University, Beijing 100871, China}
\email{xushenzhen@pku.edu.cn}

\correspondingauthor{Lile Wang, Shenzhen Xu}
\email{lilew@pku.edu.cn, xushenzhen@pku.edu.cn}

\begin{abstract}
  The adsorption of volatile molecules onto dust grain
  surfaces fundamentally influences dust-related processes,
  including condensation of gas-phase molecules, dust
  coagulation, and planet formation in protoplanetary
  disks. Using advanced {\it ab-initio} density functional
  theory with r$^2$SCAN+rVV10 van der Waals functionals, we
  calculate adsorption energies of H$_2$, H$_2$O, and CO on
  carbonaceous (graphene, amorphous carbon) and silicate
  (MgSiO$_3$) surfaces. Results reveal fundamentally
  different adsorption mechanisms: weak physisorption on
  carbonaceous surfaces
  ($|\Delta\epsilon_{\rm ad}|\sim 0.1-0.2~{\rm eV}$) versus
  strong chemisorption on silicates
  ($|\Delta\epsilon_{\rm ad}|\sim 0.5-1.5~{\rm eV}$) via
  coordination bonds. Kinetic Monte Carlo simulations
  incorporating these energies demonstrate divergent surface
  evolution: carbonaceous grains exhibit distinct
  condensation radius compared to silicates, while the
  cocrystal of H$_2$O and CO significantly increases the
  desorption temperature of CO. The actual radii of
  gas-phase molecule depletion could thus be a comprehensive
  result of temperatures, chemical compositions, and even
  evolution tracks. Meanwhile, silicates maintain
  chemisorbed molecular coatings throughout most disk
  regions. Such dichotomy in surface coverage could also
  provide a natural mechanism for carbon depletion in inner
  planetary systems.
\end{abstract}


\section{Introduction}
\label{sec:intro}

The adsorption of volatile chemical species onto the
surfaces of dust grains is of broad and critical interest to
researchers studying the interstellar medium (ISM), star
formation, and planet formation.  For example, in
protoplanetary disks (PPDs hereafter), dusts may play a
central role in controlling the thermochemical
conditions. Grains also hold the chemical equilibrium of
formation versus dissociation inside disks, as their
surfaces are the places where most \chem{H_2} and
\chem{H_2O} are formed \citep[e.g.][]{2002ApJ...575L..29C,
  2004ApJ...604..222C, 2014ApJ...786..135A}. Moreover, dust
surfaces can adsorb gas-phase charge carriers (ions and free
electrons) rather efficiently, letting these carriers
deposit their free charge and get neutralized as they get
desorbed (e.g. \citealt{2006A&A...445..205I}; see also
\citealt{Bai+Goodman2009}). This mechanism maintains a low
but non-zero fraction of ionization in the pertinent zones
in PPDs, enabling the non-ideal magnetohydrodynamic (MHD)
effects to drive disk winds and accretion flows efficiently
\citep{Bai+Goodman2009, 2011ApJ...739...51B,
  2017ApJ...845...75B, 2019ApJ...874...90W}.  Volatile
coatings on dust grain surfaces, for example water, regulate
grain sintering and thereby shape the dust size distribution
throughout PPD evolution
\citep{2016ApJ...821...82O,2017ApJ...845...31H}.  Such
modulated distributions could influence their interactions
with charged particles and lead to the formation of
sub-structures in PPDs by alternating the magnetic
diffusivities, the key parameters determining the
interactions between the magnetic fields and the gas
\citep[e.g.][]{2019ApJ...885...36H,
  2021ApJ...913..133H}. More generally, the dynamic events
related to dust growth, including coagulation, bouncing,
breaking, and shattering of grains, are sensitively
dependent on the sticking energy between surfaces
\citep{1993ApJ...407..806C, 1997ApJ...480..647D}. Such
sticking energy should also be susceptible to the conditions
on the grain surfaces. Discussions regarding these events
are extensive under various circumstances, including
theoretical analyses \citep[e.g.][]{2004ApJ...616..895Y,
  2009MNRAS.394.1061H, 2009A&A...502..845O,
  2014MNRAS.437.1636H}, and even experimental studies
\citep{2000SSRv...92..265B}.

As the crucial characterization of adsorption processes, the
energy of adsorption largely determines the status of
volatile species and dust grains themselves in dust-rich
astrophysical systems. It is widely assumed, based on
observations and experiments, that the astrophysical dust
grains can be broadly categorized into silicate grains and
carbonaceous grains, whose mass fractions are in the same
order of magnitude (see e.g. \citealt{1994A&A...292..641J,
  1995A&A...300..503D, 1998A&A...332..291J,
  2000A&A...364..282F, 2010ARA&A..48...21H}; see also the
data compilations in \citealt{DraineBook}).

Over the past few decades, the adsorption condition of key
species has been measured mainly by experimental
studies. These experimental results should, nevertheless, be
clarified and specified for their applicability and
compatibility. For example, the reported adsorption energy
of water on graphite is measured to be
$E_\ad / \kb\simeq 4800~\K$, almost identical to the
adsorption on silicates \citep{1990Icar...87..188S,
  2004MNRAS.354.1141V, 2005JChPh.122d4713B,
  2006A&A...445..205I, 2022ESC.....6..597M}. These values
neveretheless seem at odds with the quantum chemistry
studies concluding that graphite is only mildly hydrophilic
by adsorbing molecules with the van der Waals (vdW
hereafter) interactions \citep[see
also][]{PhysRevB.84.033402, hamada2012adsorption,
  brandenburg2019physisorption}, while the silicate surfaces
could adsorb water mainly by covalent or coordination bonds
and hydrogen bonds. Additionally, experimental studies have
focused on the single-species adsorption conditions only,
while the adsorbates in PPDs could be a mixture of multiple
molecular species. The impact of the surface coverage
conditions of grains deserve further and more thorough
discussions with modern theoretical and computational
methods.

In this paper, we will thoroughly study the problems related
to the adsorption of molecules via computational
chemistry. We will focus on the adsorption of water
(\chem{H_2O}) and carbon monoxide (\chem{CO}), the most
abundant molecules in the PPDs and molecular clouds (other
than \chem{H_2} and \chem{He}), onto the representative dust
grain surface models, emphasizing the physical pictures of
the results and compare them with experiments.  This paper
is structured as follows.  \S\ref{sec:micro} describes the
{\it ab-initio} calculation methods of adsorption on
carbonaceous and silicate grains, assuming single and
multiple molecule adsorption scenarios.  Using these data,
\S\ref{sec:kmc} evaluates the surface coating conditions of
both carbonaceous and silicate dust grains.
\S\ref{sec:summary} summarizes the paper, and extends the
scope of the results to the gas-phase molecule condensation
and sublimation processes onto grain surfaces, the potential
impact on planet formation mechanisms.

\begin{figure}
  \centering
  \includegraphics[width=0.35\textwidth,keepaspectratio]
  {\figdir/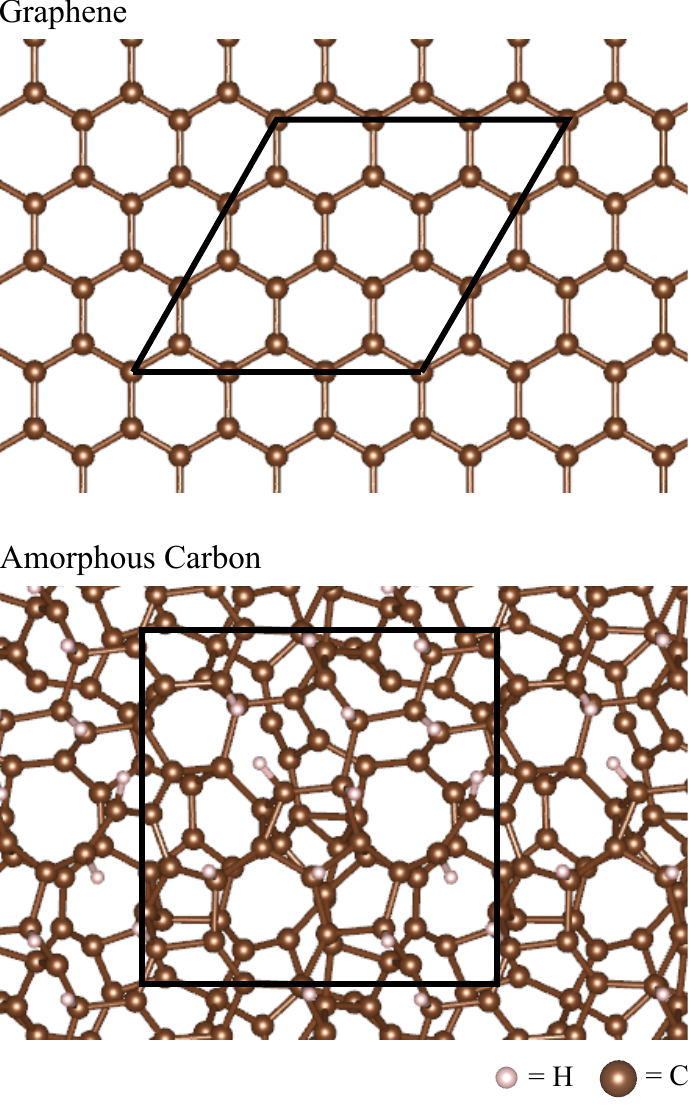}
  \caption{Top-down views of the model carbonaceous
    substrates studied in this article, including graphene
    (upper panel) and amorphous carbon (lower panel). The
    supercells used in the DFT calculations are illustrated
    by black solid lines in each panel, respectively.}
  \label{fig:carbon_base}
\end{figure}

\section{Ab-initio adsorption energy calculations}
\label{sec:micro}

\begin{figure*}
  \centering
  \includegraphics[width=0.9\textwidth, keepaspectratio]
  {\figdir/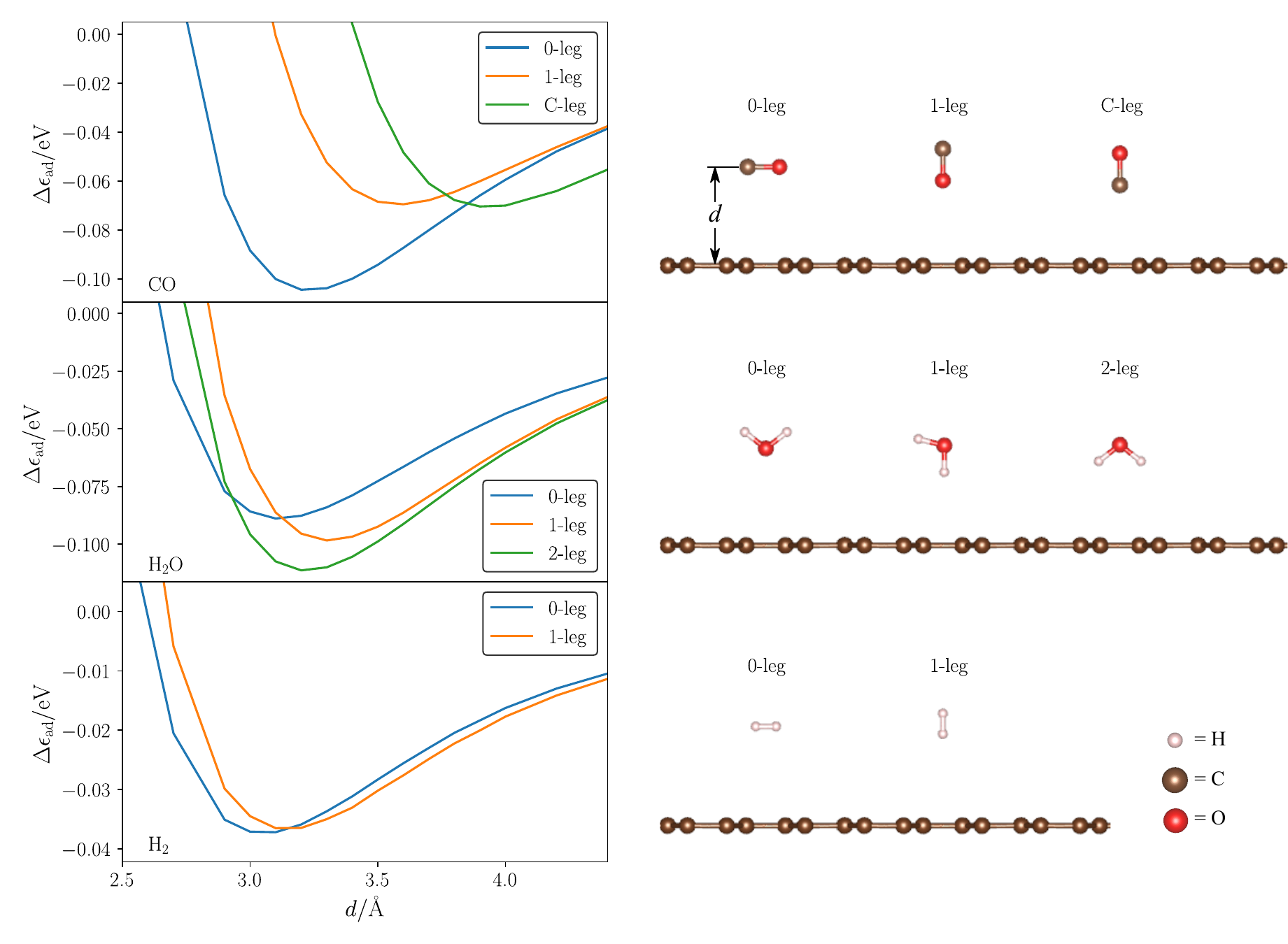}
  \caption{{\bf Left column}: Adsorption energy as functions
    of molecular orientations and distances to the
    substrates for \chem{CO} (upper row), \chem{H_2O}
    (middle row), and \chem{H_2} (lower row) molecules over
    graphene substrates. {\bf Right column}: Illustrations
    of the molecular orientation notations. The depicted
    molecules correspond to the energy curve panels in the
    left column.}
  \label{fig:e_ads_graphene}
\end{figure*}

The condensation process of key molecules onto astrophysical
grain surfaces is calculated in this work by examining
adsorption conditions using advanced computational chemistry
techniques. The \textit{ab-initio} density functional theory
(DFT) package VASP (\citealt{PhysRevB.47.558,
  PhysRevB.54.11169}) is adopted, utilizing projector
augmented wave (PAW) method calculations with PAW potentials
for the atoms involved. To ensure accurate estimations of
van der Waals (vdW) interactions, the combination of
r$^2$SCAN \citet{PhysRevLett.115.036402,
  furness2020accurate} and rVV10 \citet{PhysRevB.87.041108}
functionals is employed for electron exchange-correlation
(XC) interactions. This combination, which involves
generalized gradient approximations (GGA) and extensions to
higher-order derivatives (meta-GGA), is confirmed by various
tests and benchmarks \citep{PhysRevX.6.041005} to handle vdW
interactions under the DFT framework accurately in a wide
range of situations, especially physisorptions. For
instance, the single-molecule physisorption of water onto
graphene, which is a highly relevant case to this work, is
found to yield an adsorption energy
$|\Delta \epsilon_{\ad}| \simeq 0.096~\eV$ when using
r$^2$SCAN+rVV10, a result that is remarkably close to
fiducial quantum Monte Carlo (QMC) calculations
(\citealt{PhysRevB.84.033402,
  brandenburg2019physisorption}).  Within the PAW paradigm,
a set of plane-wave bases is utilized to expand the
electronic wave functions, with a cutoff energy of
$600~\eV$.

To measure the interaction energy of the systems, the
molecule of interest is first positioned at the desired
configuration to obtain the total energy with interaction,
$\epsilon_{\mathrm{tot}}$. The molecule $X$ is then moved
sufficiently far from the substrates to ensure that
interactions between $X$ and the substrate (as well as
neighboring adsorbates, in some cases) became
negligible. The DFT calculation is repeated to obtain the
sum of the energies of $X$ and the substrate,
$\epsilon_X + \epsilon_{\mathrm{sub}}$. Based on the data
collected, the adsorption energy for the given adsorption
configuration is subsequently defined as,
\begin{equation}
  \label{eq:e-ads}
   \Delta \epsilon_{\ad} =
   \epsilon_{\rm tot} - (\epsilon_X + \epsilon_{\rm sub})\ .
\end{equation}
In cases of adsorption, such convention yields
$\Delta\epsilon_{\ad} < 0$. It should be noted that this
adsorption energy is derived from the $0~\K$ potential
energy surface (PES) without thermodynamic corrections;
thus, the results must be adjusted for thermodynamic effects
and zero-point energy before application in
finite-temperature physical scenarios
(Appendix~\ref{sec:apdx-adsorb-therm-corr}).

\subsection{Adsorption onto Carbonaceous Surfaces}
\label{sec:ads-carbon}

Carbonaceous dust grains are considered to exist in
graphitic-like states, including graphites, oxidized
graphites, graphene, and even polycyclic aromatic
hydrocarbons (PAHs) at the smallest sizes
$\lesssim 0.01~\micron$ (see also \citealt{DraineBook}), or
as amorphous carbon. In terms of their astrophysical
properties, these materials are observed to behave
similarly, with the surface layer of carbon atoms dominating
due to the weak vdW coupling between graphitic layers.

\subsubsection{Single-molecule Adsorption onto Graphene}
\label{sec:ads-carbon-graphene}

The adsorption of single molecules onto a single layer of
graphene and amorphous carbon is first investigated as
fiducial cases in this study. Additional calculations are
conducted to verify that various situations, such as
multiple adsorbates, yield quantitatively consistent results
regarding the properties of interest. The construction of
the graphene substrate is described in
Appendix~\ref{sec:apdx-carbon-graphene}.

Beginning with water molecules, following
\citet{brandenburg2019physisorption}, the molecules are
positioned above the center of a carbon hexagon, with
orientations classified as ``0-leg'', ``1-leg'', and
``2-leg'' (illustrated in
Figure~\ref{fig:e_ads_graphene}). For each orientation, a
series of $\Delta E_{\rm abs}$ values is obtained at
different displacements $d$ from the substrate surface,
measured from the plane of carbon nuclei to the
center-of-mass of the adsorbate molecule. It should be noted
that during this step, the ionic structure of the system is
fixed to manually iterate through configuration space. This
setup ensured that the adsorbate configuration—particularly
its orientation and distance to the substrates—remained
unchanged, allowing correct energy curves to be
obtained. The resulting energy curves for these
configurations are presented in
Figure~\ref{fig:e_ads_graphene}.  The difference between our
approach (the r$^2$SCAN + rVV10 method) and the QMC method
for the 1-leg case \citep{PhysRevB.84.033402} is
$\lesssim 0.05~\eV$, well within the range of variations
among different XC methods \citep{hamada2012adsorption}.
Similar calculations for \chem{H_2} and \chem{CO} are
performed following the same paradigm, with molecular
orientation notations illustrated in
Figure~\ref{fig:e_ads_graphene}. For completeness and
verification purposes, the adsorption of helium is also
examined, yielding an adsorption energy of
$|\Delta\epsilon_{\ad}| \lesssim 0.07~\eV$.

\begin{figure}
  \centering
  \includegraphics[width=0.49\textwidth, keepaspectratio]
  {\figdir/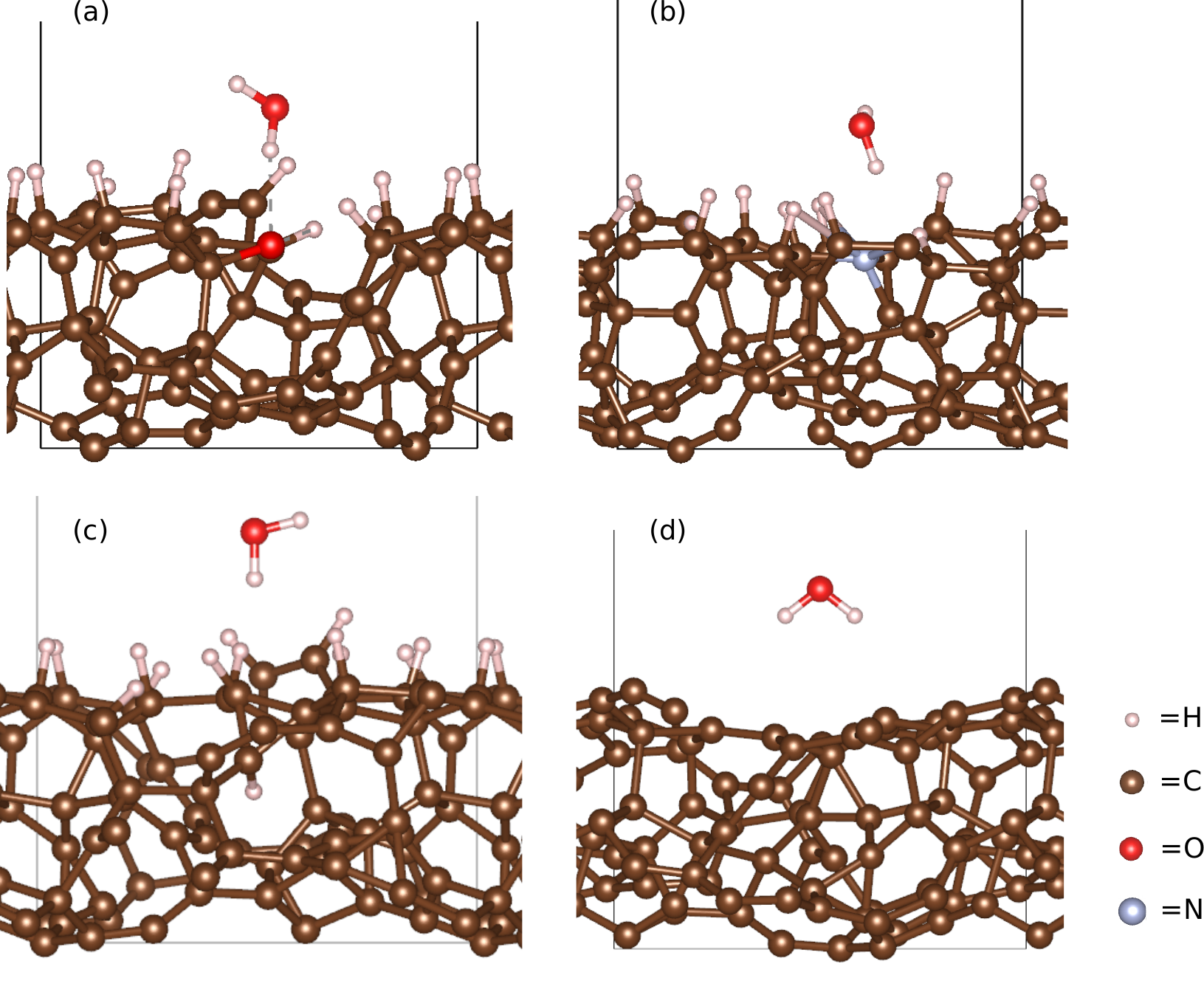}
  \caption{Example of \chem{H_2O} single-molecule
      adsorption onto carbonaceous surface under various
      situations, placed above: (a) ``doped'' oxygen atom,
      (b) doped nitrogen atom, (c) hydrogen confined in the
      bulk substrate, and (d) surface not covered by
      hydrogen. }
  \label{fig:e_ads_var}
\end{figure} 

\begin{deluxetable*}{lcccccc}
  \tablecolumns{7}   
  % \tabletypesize{\scriptsize}
  \setlength{\tabcolsep}{10pt}      
  \tablecaption{Physical parameters of adsorbates on
    amorphous carbon grain
    surfaces. \label{table:ads-nbr-carbon} } 
  \tablehead{
    \colhead{Adsorbates} & \colhead{Neighbors} &
    \colhead{$\nu/{\rm THz}^\dagger$}
    & \colhead{$\xi/{\rm THz}^\dagger$} &
    \multicolumn{3}{c}{$|\Delta
      \epsilon_{\ad}|/\eV$ with neighbors}
    \\
    \cline{5-7}
    \colhead{} & \colhead{} & \colhead{} & \colhead{} &    
    \colhead{$0$} & \colhead{$1$} & \colhead{$\max^\ddagger$} 
  } 
  \startdata
  \chem{H_2O} & \chem{H_2O} & $\sim 3.5$ & $\sim 2.0$ & 0.16
  & 0.36 & 0.87 \\
  & \chem{CO}   & & & & 0.26 & 0.35 \\ 
  \cline{2-7}
  \chem{CO} & \chem{CO}   & $\sim 3.6$ & $\sim 3.0$ & 0.10 &
  0.13 & 0.14 \\ 
  & \chem{H_2O} &             &            &      & 0.24 & 0.27 \\
  \cline{2-7}
  \chem{H_2}$^*$ & \chem{H_2} & $\sim 5.0$ & $\sim 3.5$ &
  0.028 & 0.029 & 0.029 \\ 
  \enddata
  \tablecomments{
    $\dagger$: The eigen frequencies of
    vertical ($\nu$) and horizontal ($\xi$) vibrations of
    adsorbates on the adsorption sites, estimated by finite
    difference
    derivatives.\\
    $\ddagger$: The maximum number of neighbors on the
    surface is generally 4
    nearest and 4 second-nearest.\\
    *: Interactions between \chem{H_2} and other types of
    adsorbates are similarly weak and not presented.
  }
\end{deluxetable*}

\subsubsection{Adsorption onto amorphous carbon surfaces}
\label{sec:ads-carbon}

The realistic composition of interstellar and protostellar
carbonaceous grains is considered to differ considerably
from graphites or graphene. According to experimental
studies \citet{1998A&A...332..291J}, these grains are more
likely to be amorphous carbon, which in various cases may be
doped with oxygen and hydrogen at approximately $\sim 1~\%$
of atomic concentration. Hydrogen atoms are expected to be
located predominantly at grain surfaces due to their
inability to achieve coordination numbers greater than
unity. Oxygen atoms may exist in both the bulk substrate and
near the surface; however, numerical experiments indicate
that surface oxygen atoms are always saturated by capturing
hydrogen atoms because of intense dipole moments. In
practice, while the cases with doped atoms are also
calculated, only hydrogen and carbon atoms are elaborated
and explored in details. The construction of amorphous
carbon substrates is explained in
Appendix~\ref{sec:apdx-carbon-amorphous}.

Numerical experiments for single-molecule adsorption energy
are conducted above the amorphous carbon substrate. The
adsorbate is positioned at five different horizontal
locations above the surface, with offsets from the
horizontal center of the substrate of transverse size
$(9.09~\ang)^2$, as defined by:
\begin{equation}
  \label{eq:single-mol-loc}
  \begin{split}
    (\Delta x,\ \Delta y)
    & \in \{
    (-1, -1), (-1, 1), (1,-1), (1,1), (0,0) 
      \} \\
    & \times \left( \dfrac{9.09~\ang}{4}, 
      \dfrac{9.09~\ang}{4} \right)\ .
  \end{split}
\end{equation}
At each site, ionic structures are relaxed to locate the
energy minimum before the adsorbate is moved sufficiently
far from the substrate to evaluate $\Delta\epsilon_{\ad}$
using equation~\eqref{eq:e-ads}. Single-molecule adsorption
energies in Table~\ref{table:ads-nbr-carbon} are obtained by
averaging $\Delta\epsilon_{\ad}$ over the five
representative sampling sites. It is acknowledged that the
inherent stochasticity and structural heterogeneity of
amorphous carbon surfaces could be better represented by a
larger ensemble of sampling points characterizing such
disordered media. Nonetheless, an exhaustive sampling of the
near-infinite configuration space of local adsorption sites
is computationally prohibitive. The amorphous carbon
substrate model generated for this study
(Appendix~\ref{sec:apdx-carbon-amorphous}) also exhibits
some degree of structural homogeneity within the supercell
that facilitates representative sampling of the dominant
binding environments. In fact, supplementary DFT
calculations have also been performed on a wider set of
prospective sites and alternative amorphous carbon
realizations \response{(see
  Appendix~\ref{sec:apdx-amorph-ene-rand} and
  Figure~\ref{fig:amorph-ene-sample})}, whose consequent
adsorption energies consistently fall within the range
established by the five primary sampling points (except a
few extreme cases, where the adsorbates eventually reside in
the sites whose $|\Delta \epsilon_{\rm ad}|$ still stay in
the consistent range after structure relaxation).

In It is noted that replacing approximately $\sim 2~\%$ of
carbon atoms with oxygen yielded nearly identical results,
as oxygen atoms are rarely exposed directly (due to hydrogen
surface coverage) and thus contributed minimally to
interactions with adsorbates. Through deliberate exposure of
oxygen and nitrogen atoms at the surface (e.g.,
Figure~\ref{fig:e_ads_var}), the adsorption energy of
\chem{H_2O} via hydrogen bonds is measured to be $0.35~\eV$
(above oxygen) and $0.29~\eV$ (above nitrogen),
respectively. The adsorption energy of \chem{CO} near doping
atoms exhibited no significant variations.

When substrate surface coverage fractions become
non-negligible, interactions between adsorbates may play a
crucial role in modifying their kinetic behaviors. The vdW
interactions and hydrogen bonds between adjacent molecules
increase the barriers for desorption and hopping, with
intensity dependent on both the numbers and types of
neighbors. These effects are quantified through numerical
experiments based on amorphous carbon substrates. The
interaction energy is measured through the following
procedures: % (1) the adsorbate and its neighbors are placed
% above the substrate, and ionic configurations are evolved
% with DFT until stress relaxation, after which the total
% energy of the relaxed system is evaluated; (2)
\begin{enumerate}
\item Initial site pre-optimization: The ``primary''
  adsorbate of interest is first placed and relaxed at a
  previously descreibed single-molecule adsorption site via
  full geometry optimization.
\item Initial neighbor positioning: Neighboring molecules
  are introduced at radial distances approximating the
  characteristic intermolecular separations found in their
  respective bulk solid phases (e.g., the hydrogen-bonding
  network of amorphous solid water when the primary molecule
  and its neighbors are all \chem{H_2O}). The azimuthal
  angles of these neighbors are selected so that the
  neighbors are close to their likely single-molecule
  adsorption site if possible (e.g., the nearest available
  hexagonal centers or ``hollow'' sites over the amorphous
  surface), or in random otherwise (while the separations
  between neighbors are still guaranteed).
 \item Constrained adsorbate relaxation: DFT structural
   optimization is conducted where the degrees of freedom of
   all adsorbed molecules are relaxed to a local energy
   minimum. During this step, the coordinates of the
   substrate atoms are held fixed to maintain the
   pre-equilibrated surface morphology.
\item Adsorption energy measurement: Adsorbate of interest
  is moved sufficiently far from the substrate
  ($\Delta z \gtrsim 10~\ang$), and configurations of all
  remaining atoms are relaxed to re-evaluate the total
  energy.
\end{enumerate}
Energy differences measured through these steps yielded the
``neighbor-enhanced'' total binding energy. For tests
involving adsorbates with one neighbor, adsorbate pairs are
positioned at five different sites (with centers of mass
offsets following equation~\ref{eq:single-mol-loc}) to
obtain average enhanced binding energies. The overall
binding energy (including adsorption) is observed to
increase sub-linearly, as vdW forces exhibit saturation
behavior and hydrogen bonds require specific molecular
orientations. Tests are also conducted to determine the
numbers of different neighbors that maximize absolute
binding energy values, beyond which binding energy decreases
due to repulsive forces. Similar to the single-molecule
cases, depleting the possibilities over the heterogeneous
amorphous carbon surfaces is virtually not
practical. Nonetheless, because of the qualitative
homogeneity of the model amorphous surface, it is expected
that such procedures could still reasonably sample the
typical situations that are relevant in the following
calculations.

\subsection{Adsorption onto Silicates}
\label{sec:silicate}

\begin{figure*}
  \centering
  \includegraphics[width=0.9\textwidth, keepaspectratio]
  {\figdir/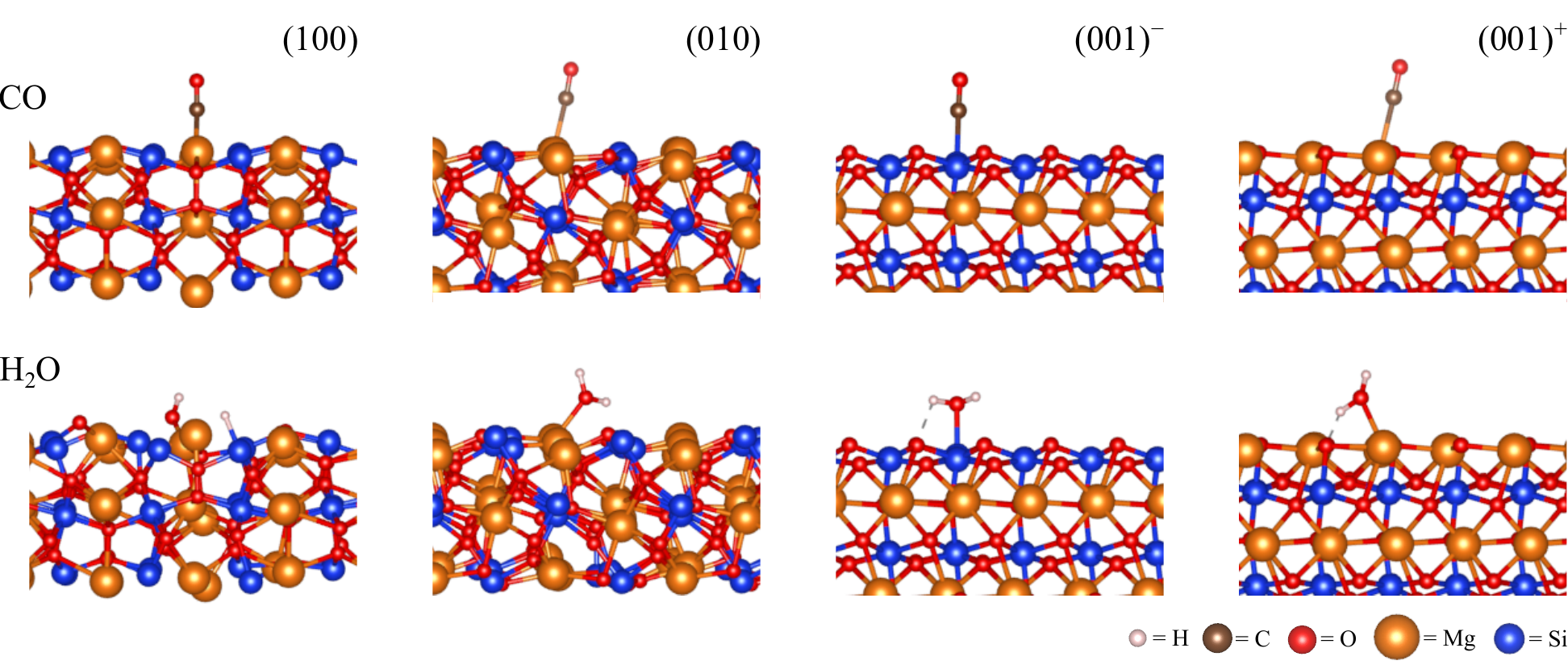}
  \caption{Configurations of adsorbed \chem{CO} molecules
    (upper row) and \chem{H_2O} molecules (lower row) over
    different crystal surfaces (indicated above each column)
    of the model silicate \chem{MgSiO_3}. It should be noted
    that adsorption of \chem{H_2O} over the $(100)$ surface
    results in molecular dissociation.}
  \label{fig:silicate_ads}
\end{figure*}

In comparison to carbonaceous grains, silicate compositions
are recognized to be considerably more complex. To maintain
model tractability while preserving physical relevance,
iron-poor enstatite \chem{MgSiO_3} (Pnma space group using
the Materials Project data, see also \citealt{MatProj}) is
selected as a proxy for magnesium silicate substrates. This
choice is motivated by the abundance of magnesium, which
among alkali and alkaline earth metals in the ISM and PPDs,
typically matches silicon abundance levels.%  To emulate
% astronomical silicate formation conditions, hydrogen atoms
% are added to exposed oxygen atoms on model silicate
% surfaces, following procedures analogous to those employed
% for amorphous carbon.
The details of silicate substrate model construction are
explained in Appendix~\ref{sec:apdx-carbon-silicate}.

Admittedly, the model crystalline Pnma enstatite is unable
to fully represent the amorphous enstatite, a commonly
encountered phase of astrosilicates also in PPDs
\citep[see][and references therein]{2010ARA&A..48...21H}.
Practical representation of disordered systems in the
ab-initio DFT frameworks, nevertheless, typically involves
significant computational trade-offs. While crystalline
substrates can be accurately modeled with relatively compact
supercells ($\sim 10^2$ atoms), a physically realistic
amorphous structure requires substantially larger,
disordered models (typically $\sim 10^3$ atoms) to minimize
periodic artifacts, making systematic potential energy
surface mapping with advanced metaGGA and van der Waals
functionals (e.g., r$^2$SCAN+rVV10 adopted in this work)
prohibitively expensive.

In this study, crystalline enstatite is adopted as a generic
proxy for its amorphous counterpart, becaus (1) both phases
share very similar stoichiometry, providing a chemically
representative building block for magnesium-rich silicates;
and (2) surface reactivity is primarily governed by local
coordination environments rather than long-range order, with
adsorption dominated by undercoordinated Mg and Si sites
whose chemical interactions remain consistent across
phases. To better approximate the morphological and chemical
heterogeneity of a disordered surface within these
computational constraints, calculations were performed
across multiple distinct facets including (100), (010), and
$(001)^{\pm}$, which expose varying ratios of Mg and Si
atoms in diverse coordination states, effectively bracketing
the range of local environments encountered on a realistic
silicate surface. 

The specific atomic composition of silicate substrates
beneath adsorbate molecules is found to determine adsorption
properties. Initial investigations positioned molecules with
varying orientations at different heights above different
atomic sites. After identifying configurations with
approximately minimal energy, molecules are placed
accordingly on the substrate, and ionic structures are
evolved to achieve relaxation, eliminating total stress to
the desired accuracy ($\lesssim 10^{-2}~\eV~\ang^{-1}$) and
thereby locating energy extrema. In astrophysical conditions
relevant to PPDs, it is possible that the silicate surfaces
could be covered by chemisorbed hydrogen. Additional tests
were performed on surfaces partially covered with these
hydrogen (not shown in this paper), which preferentially
saturates exposed under-coordinated oxygen atoms while
leaving some Mg and Si coordination centers
accessible. These tests indicate that hydrogen coverage does
not significantly alter the adsorption energies of
\chem{H_2O} and CO on enstatite surfaces (with changes
limited to $\lesssim 0.2~\eV$), which is of secondary
importance compared to the total chemisorption energy of
$|\Delta \epsilon_{\rm ad}|\gtrsim 1~\eV$. This is
consistent with expectation, as non-closest-neighbor
conditions do not strongly affect covalent coordinate bonds.

Configurations converged at energy extrema are illustrated
in Figure~\ref{fig:silicate_ads}, with corresponding
adsorption energies presented in
Table~\ref{table:ads-silicate}. The adsorption energy values
for \chem{H_2O} and \chem{CO} are generally observed to be
one order of magnitude greater than those on carbonaceous
surfaces, clearly indicating chemisorption processes. This
enhancement is attributed to the coordination numbers of
metal centers (Si atoms on $(001)^-$ and Mg atoms elsewhere)
being significantly less than the saturation value of 6 in
bulk \chem{MgSiO_3}, thereby permitting coordination bond
formation and elevating adsorption energies to approximately
$\sim 1~\eV$. A notable exception is identified for
\chem{H_2O} adsorption on the (100) surface where the
molecule adsorption eventuall leads to dissociation.  On
this facet, exposed magnesium atoms exhibit a preferred
directionality for coordination bond formation that lies
nearly tangential to the surface plane, positioning the
chemisorbed water molecule close to neighboring
under-coordinated silicon atoms (with coordination numbers
of $\sim 4$, versus 6 in the bulk). These Si sites could
accept and stabilize the transferred proton, collectively
lowering the activation barrier for O--H bond cleavage. In
comparison, the (010), $(001)^+$, and $(001)^-$ surfaces
favor coordination geometries that extend more
perpendicularly from the surface, keeping sufficient
distances of the water molecule from under-coordinated
atoms, while exposing more fully coordinated Si--O--Si
bridging configurations that are less susceptible to
protonation.
% exceptionally high adsorption energy
% resulted in molecular dissociation.

Chemical bonding is observed to constrain molecular
orientations following convergence. In \chem{H_2O}, the
oxygen atom functions as the electron donor, while in
\chem{CO}, the carbon atom serves this role. Given the
sufficiently high band intensities and their strong
orientation dependence, numerical experiments confirmed that
vdW or hydrogen bond interactions between adjacent
adsorbates could be neglected on silicate substrates. The
increased elasticity resulting from stronger bonds also
produced higher eigen frequencies of molecular oscillation
($\nu$ and $\xi$). In contrast, the inability of \chem{H_2}
to form coordination bonds resulted in negligible adsorption
energy.

\begin{deluxetable*}{lcccccc}
  \tablecolumns{7}   
  % \tabletypesize{\scriptsize}
  \setlength{\tabcolsep}{10pt}      
  \tablecaption{Physical parameters of adsorbates on
    different surfaces of model silicate \chem{MgSiO_3}.
    \label{table:ads-silicate}}
  \tablehead{
    \colhead{Adsorbates} &
    \colhead{$\nu/{\rm THz}$}
    & \colhead{$\xi/{\rm THz}$} &
    \multicolumn{4}{c}{$|\Delta \epsilon_{\ad}|/\eV$
      on surfaces }
    \\
    \cline{4-7}
    \colhead{} & \colhead{} & \colhead{} &
    \colhead{$(100)$} & \colhead{$(010)$} &
    \colhead{$(001)^-$} & \colhead{$(001)^+$}
  }
  \startdata
  \chem{H_2O} & $\sim 13$ & $\sim 6.0$
  & $-^{**}$ & $1.50$ & $1.32$ & $1.17$ \\
  \chem{CO} & $\sim 6.7$ & $\sim 5.5$
  & $1.23$ & $0.95$ & $0.74$ & $0.45$ \\
  \chem{H_2} & $\sim 6.3$ & $\sim 3.0$
  & $0.015$ & $0.008$ & $0.016$ & $0.025$ 
  \enddata
  \tablecomments{Similar to
    Table~\ref{table:ads-nbr-carbon}, but the neighbor
    conditions are not included as the molecules are
    chemisorbed onto silicate surfaces (see also
    \S\ref{sec:silicate}).
    \\
    $**$: Molecule dissociated after chemisorption.}
\end{deluxetable*}

\section{Kinetic Monte Carlo Simulations of Grain Surfaces}
\label{sec:kmc}

Surface coverage conditions of dust grains play a
fundamental role in regulating coagulation processes through
their influence on sticking behavior and surface energy
dynamics. To systematically investigate these coverage
conditions, we developed a specialized kinetic Monte Carlo
(KMC) simulation code, optimized for GPU acceleration to
handle the computational demands of these complex
simulations\footnote{\url{https://github.com/wll745881210/kmc_ads}.}
whose general procedures are explained in
Appendix~\ref{sec:apdx-kmc-method}.  The zero-temperature
results obtained in \S\ref{sec:micro} need thermodynamic
corrections before being applied in the KMC simulations (see
Appendix~\ref{sec:apdx-adsorb-therm-corr}). It is also
noticed that the KMC approach itself inherently sacrifices
the accuracy of ab-initio calculations by discretizing
events and transition rates on a square lattice, rather than
employing rigorous off-lattice simulations that directly
utilize consistent DFT calculations. In this work, the
transition rates are also approximate values derived from
limited sampling of substrate conditions, based on
adsorption and binding energies obtained from a limited-size
amorphous carbon model or from crystalline enstatite as a
proxy for amorphous silicate. Nevertheless, the purpose of
conducting KMC simulations is to balance this trade-off, to
obtain improved statistical sampling over larger systems at
the cost of single-process accuracy. By using test results
based on a model amorphous carbon surface with reasonable
homogeneity, and by bracketing amorphous silicate surface
environments with multiple crystalline facets, the KMC
results are expected to retain reasonable semi-quantitative
reliability.

\subsection{Setups of KMC simulations}
\label{sec:kmc-method}

The KMC simulations are implemented on two-dimensional
periodic square lattices designed to emulate grain surfaces,
where each lattice point corresponds to a discrete
adsorption site. This discretization approach allows for
efficient modeling of molecular adsorption, desorption, and
migration processes across the surface landscape.  The KMC
framework incorporated both square and hexagonal lattice
geometries to assess potential structural dependencies in
surface coverage simulations. Comparative analyses revealed
that hexagonal lattices produced nearly identical results to
square lattices in terms of surface coverage conditions,
indicating that the fundamental physical processes governing
molecular adsorption exhibit minimal sensitivity to lattice
symmetry. Consequently, for enhanced computational
efficiency and analytical clarity, only square lattice
implementations are retained for detailed presentation and
comparative analyses throughout this investigation.

Specific initial and terminal conditions are established to
ensure physical relevance and numerical stability. Each
simulation commenced with a pre-defined ``condensation
nucleus'' consisting of a $4\times 4$ patch of water
molecules positioned on the surface. This initialization
strategy prevents numerical artifacts associated with
supercooling or superheating conditions that could yield
unreliable results. Simulation convergence is defined by
either the complete disappearance of the condensation
nucleus or the achievement of surface coverage exceeding
$99\%$. In cases of extensive coverage, where second and
higher molecular layers begin to form, analyses are
deliberately restricted to the first monolayer to maintain
interpretational clarity.

\subsection{Surface coverage conditions simulated by KMC}
\label{sec:kmc-results}

The divergent surface coating behaviors observed in
protoplanetary disks (PPDs) directly originate from the
fundamentally different adsorption physics governing
carbonaceous and silicate grains. Within a representative
PPD model approximating early Solar System conditions, we
conduct kinetic Monte Carlo simulations based on vdW-DFT
computational results. These KMC implementations
comprehensively account for factors influencing desorption
rates, including finite-temperature thermodynamic
corrections and inter-adsorbate interactions parameterized
through the eigen frequencies of adsorbate vibrations
(Tables~\ref{table:ads-nbr-carbon} and
\ref{table:ads-silicate}).

Simulations are conducted for dust grains situated in the
mid-plane of a model PPD, employing density and temperature
profiles consistent with a young stellar object of
bolometric luminosity $L\sim 4L_\odot$. The temperature
profile follows $T\propto R^{-0.4}$ due to surface flaring
effects \citet{1997ApJ...490..368C}, with specific forms
given by:
\begin{equation}
  \label{eq:disk-profile}
  \begin{split}
      & T \simeq 280~\K \left( \dfrac{R}{\au}
      \right)^{-0.4}, \\
      & \rho_0 \simeq 3.8 \times 10^{-10}~\g~\cm^{-3}
      \left( \dfrac{R}{\au} \right)^{-2.8}.
  \end{split}
\end{equation}
This mid-plane density corresponds to a surface density
profile of
$\Sigma \simeq 600~\g~\cm^{-1.5}
(R/\au)^{-1.5}$. Molecular number densities are
derived assuming standard abundance ratios:
$X(\chem{H_2})\simeq 1$,
$X(\chem{CO})\simeq 1.8\times 10^{-4}$, and
$X(\chem{H_2O})\simeq 1.4\times 10^{-4}$. Helium atoms are
excluded from consideration due to their negligible
adsorption efficiency on grain surfaces.

\begin{figure*}
  \centering
  \includegraphics[width=0.6\textwidth,keepaspectratio]
  {\figdir/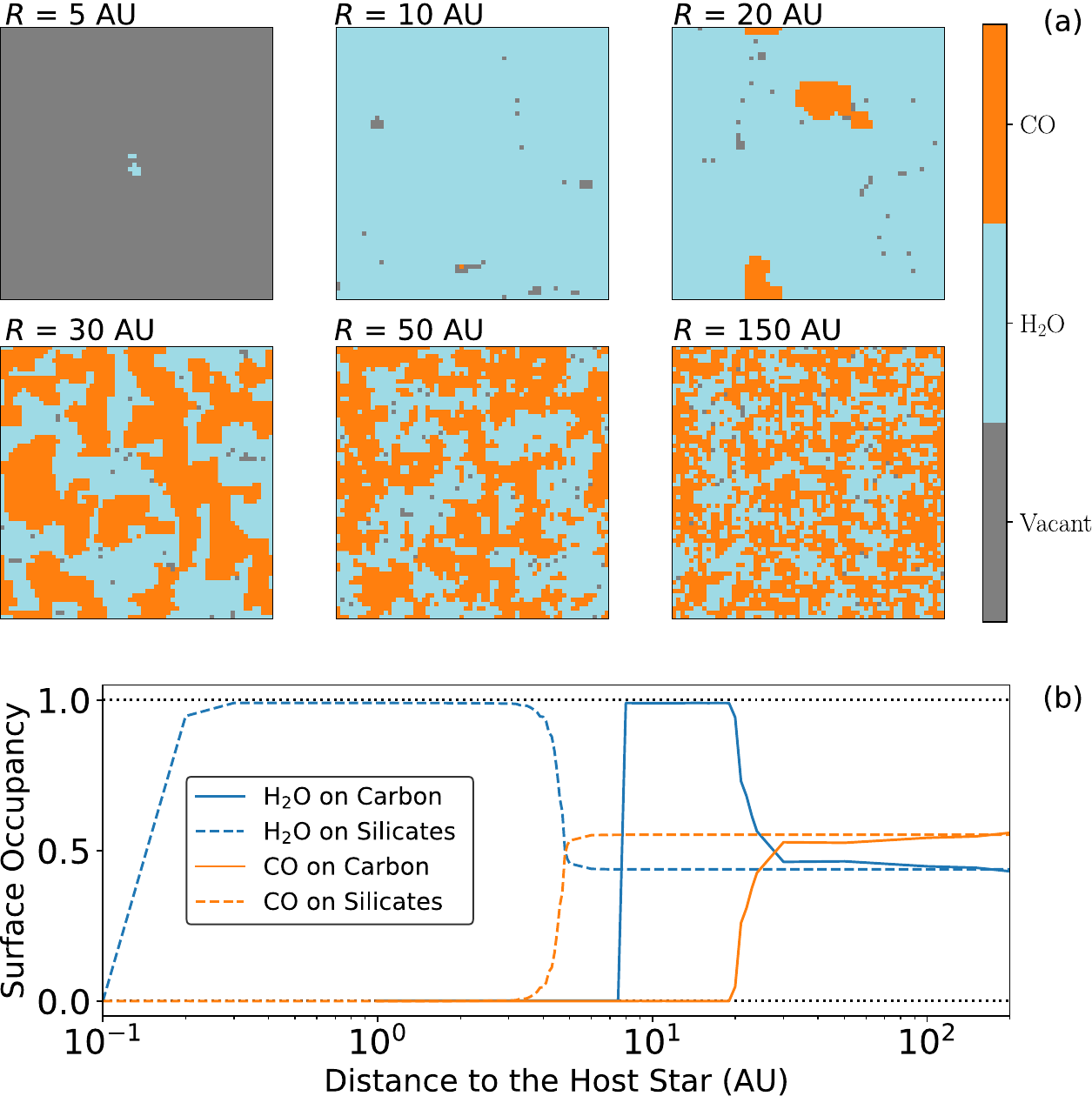} 
  \caption{ Surface coverage conditions of grains in the
    mid-plane of a Solar-system-equivalent PPD model. Panel
    (a) illustrates the coating conditions of amorphous
    carbon grains given by converged KMC calculations,
    showing a sample surface with $64\times 64$ site (each
    site separated by $3~\ang$, the typical separation
    between adsorption sites). Panel (b) plots the surface
    occupancy conditions for amorphous carbon and the
    $(001)^+$ surface of \chem{MgSiO_3}. }
  \label{fig:surface_coverage}
\end{figure*}

Across the radial domain of $0.1 < (R/\au) < 200$, KMC
simulations revealed distinct surface coverage patterns for
carbonaceous surfaces and the $(001)^-$ silicate surface, as
illustrated in Figures~\ref{fig:surface_coverage}. Silicate
surfaces undergo a transition from bare to coated states at
$R \lesssim 0.2~\au$, a consequence of the substantial
chemisorption energies characteristic of \chem{CO} and
\chem{H_2O} on these substrates. In contrast, KMC results
clearly identify a snowline on grain surfaces for water on
carbonaceous grains at $R\simeq 8~\au$ ($T\simeq 120~\K$),
interior to which surfaces remain uncoated due to elevated
thermal energies (Figure~\ref{fig:surface_coverage}). This
snowline represents a distinct concept from the conventional
snowline marking \chem{H_2O} ice condensation at
$T\sim 170~\K$ \citep{2000ApJ...528..995S}, with the former
being more directly relevant to substrate material
coagulation processes. Employing the previously assumed
adsorption energy of $\simeq 0.45~\eV$ would reposition the
\chem{H_2O} snowline at $R\simeq 2~\au$.

The fundamental differences in adsorption behavior between
carbonaceous and silicate substrates have likely remained
obscured in previous research due to systematic
misinterpretations of temperature-programmed desorption
(TPD) experimental data. This misinterpretation stems from
the complex nature of water cluster formation on substrate
surfaces. Within water molecule clusters, each molecule
typically participates in two hydrogen bonds on average,
mirroring the structural configuration of hexagonal ice
crystals (I$_{\mathrm{h}}$). This hydrogen bonding network
contributes an internal binding energy of approximately
$\epsilon_{\mathrm{hb}}\sim 0.42~\mathrm{eV}$ per molecule
(where the subscript ``hb'' denotes hydrogen bonds; see also
\citealt{silverstein2000strength}).

Conventional TPD experiments are conducted under conditions
substantially different from astrophysical environments:
temperatures well below the triple point of water and
gas-phase molecule number densities orders of magnitude
higher than those found in protoplanetary disks or the
interstellar medium. These experimental conditions promote
the formation of extensive water clusters on substrate
surfaces \citet{PhysRevB.79.235440, PhysRevB.84.033402,
  bjorneholm2016water}. Consequently, molecular addition or
removal processes during TPD measurements primarily occur at
the interfaces of these water clusters rather than at the
actual substrate-adsorbate boundary. This clustering
phenomenon could lead to a possible misinterpretation of TPD
data: the slopes observed in Arrhenius plots actually
reflect $\epsilon_{\mathrm{hb}} / \kb$, the energy scale
associated with hydrogen bond breaking, rather than the true
substrate adsorption energy $\Delta\epsilon_{\ad} /
\kb$.  For silicate substrates, the situation is
further complicated by the exceptional strength of
chemisorption bonds, which remain intact at typical
protoplanetary disk temperatures ($T\lesssim
10^3~\K$). Under these conditions, any molecular desorption
processes are governed by hydrogen bond rupture within water
clusters, again causing TPD measurements to yield
$\epsilon_{\mathrm{hb}}$ rather than the fundamental
substrate adsorption energy.
\response{Similar phenomena have also been observed in
  recent experiments of other types of molecules
  (e.g. \citealt{2017ApJ...837...65H} for \chem{CO_2}).
}
Accurate interpretation of
adsorption mechanisms and binding energies is therefore
essential for reliable modeling of astrophysical processes
involving dust grains.

\section{Discussions}
\label{sec:discussions}

We have combined vdW-corrected \textit{ab-initio} DFT
calculations and kinetic Monte Carlo simulations to quantify
how key volatiles (\chem{H_2O}, \chem{CO}, and \chem{H_2})
adsorb onto carbonaceous and silicate grain surfaces, and
how these material-dependent adsorption energies, combined
with KMC simulations, translate into different surface
coverage patterns. The results reveal a clear dichotomy
between weak physisorption on carbonaceous grains and strong
chemisorption on silicates, with important consequences for
grain growth and volatile partitioning. We notice that,
although crystalline enstatite is employed as a proxy for
more complex amorphous silicates, the fact that
chemisorption energies on silicate surfaces are an order of
magnitude stronger than on carbonaceous surfaces remains
relatively robust: the adsorption energy magnitudes
($|\Delta \epsilon_{\rm ad}| \gtrsim 1~\eV$ for \chem{H_2O}
and $\sim 0.5-1~\eV$ for CO) are not significantly
influenced by the crystalline nature of the substrate
(Tables~\ref{table:ads-nbr-carbon} and
\ref{table:ads-silicate}).

This section places these findings in a broader
astrophysical context.  We examine how history-dependent
initial surface conditions on grains can shift molecular
snowline locations (\S\ref{sec:kmc-init-cond}), investigate
how multi-species adsorption, particularly CO--\chem{H_2O}
interactions, modifies CO surface snowlines and gas-phase
depletion (\S\ref{sec:kmc-cocrystal}), and connect these
processes to molecular condensation and gas-phase depletion,
contrasting carbon abundances in different planetary
populations, and the ingredients that should be incorporated
into future models of dust evolution and planet formation
(\S\ref{sec:co-condensation}).

\subsection{Dependence of snowline locations on \\
  evolutionary history}
\label{sec:kmc-init-cond}

The initial surface conditions of dust grains could be
crucial in determining the locations and characteristics of
molecular snowlines in protoplanetary disks. When grain
surfaces begin with complete \chem{H_2O} molecular coatings
rather than initially bare configurations, the adsorbed
water molecules demonstrate significantly enhanced thermal
stability due to pre-existing intermolecular binding
networks. This phenomenon creates metastable surface
conditions that share fundamental physical characteristics
with superheating effects observed in other first-order
phase transitions.

The enhanced thermal stability arises from the collective
binding energy provided by hydrogen bonding networks between
adjacent water molecules. These networks create energy
barriers that substantially exceed the single-molecule
adsorption energy, allowing molecular layers to persist at
temperatures far above those predicted by simple adsorption
models. Consequently, the ultimate surface coverage
conditions and associated gas-phase condensation processes
(including potential molecular depletion) become strongly
dependent on the specific thermal and dynamical evolution
history experienced by dust grains.

The divergence between different evolutionary pathways can
be dramatic. When pre-hydrated grains undergo gradual
heating from cold initial conditions, they may maintain
surface coatings up to temperatures approximately a few tens
of Kelvin (typically $\sim 20-30~\K$) higher than the
equivalent snowline temperature for condensation onto
initially bare surfaces (not shown in this paper). This
effect arises because the formation of new ice layers
requires nucleation events that possess significant
activation energies, while the removal of existing layers
requires overcoming the collective binding energy of the
entire molecular network. Therefore, accurate and consistent
determination of snowline locations and condensation
conditions necessitates comprehensive modeling that
incorporates the full thermodynamic history of disk
material, including radial drift trajectories and temporal
temperature variations. The assumption of instantaneous
equilibrium between adsorption and desorption processes
proves insufficient for capturing these history-dependent
effects, requiring instead the implementation of kinetic
models that track the evolutionary pathways of individual
grain populations throughout their disk residence.

The final surface coverage of CO on dust grains is not a
static property but depends on the specific thermal and
dynamical evolution path of the dust population. The initial
conditions of grains, such as starting as ``dry grains''
that are too cold to have undergone significant thermal
processing, can preset the availability of binding sites and
the efficiency of subsequent ice mantle
formation. Furthermore, the radial transport of grains plays
a crucial role. Icy grains drifting inward from the outer
disk sublimate upon crossing their respective snowlines,
temporarily enhancing the local gas-phase abundance before
potentially re-condensing onto other grain surfaces under
the new thermal conditions.

Consequently, history-dependent snowline locations could
vary in different disks, occurring at radii corresponding to
different temperatures. Simply inferring their locations
based on a fixed temperature (e.g., $20~\K$ for CO) may
therefore lead to mismatches. Dust rings and gaps have been
found to be nearly ubiquitous in disks. The prevalent
hypothesis for their origin is that they are caused by
planet–disk interactions
\citep{dong15gap,dong17doublegap,dong17gap}, while the
snowline hypothesis \citep{zhang15} has been questioned due
to mismatches between the observed ring/gap locations and
the expected snowline locations
\citep{long18,bae18,vandermarel19}. Re-examining the
connection between the two, while accounting for each disk's
thermal and dynamical history, is therefore needed.

\subsection{Impact of multi-species adsorption on CO surface
  snowlines and gas depletion}
\label{sec:kmc-cocrystal}

The analysis of surface coverage conditions reveals that,
based on the disk temperature and hydrodynamic profiles,
$R\sim 20~\au$ approximately divides the grain surface
snowline for \chem{CO} on amorphous carbon grains,
corresponding to a local temperature of $T\simeq 80~\K$
according to the disk temperature profile defined in
equation~\eqref{eq:disk-profile}. This temperature appears
anomalously high when compared to existing experimental and
computational studies, which typically report CO
condensation temperatures around $\sim 20~\K$
\citep{2017ApJ...849...80C}. However, detailed examination
of Figure~\ref{fig:surface_coverage} indicates that CO
condensation initiation is not an isolated process but
rather occurs through coordinated condensation with
\chem{H_2O} molecules.

The underlying mechanism for this elevated condensation
temperature becomes evident upon inspection of
Table~\ref{table:ads-nbr-carbon}, which demonstrates that
the interaction energy between a \chem{CO} molecule and an
adjacent \chem{H_2O} molecule significantly exceeds that
between two \chem{CO} molecules. This enhanced
intermolecular binding substantially increases the effective
adsorption energy of CO molecules when they are incorporated
within a water ice matrix. The resulting configuration
resembles ``cocrystal'' structures, where CO molecules
become trapped within the lattice of water ice (distinct
from traditional clathrate hydrates, particularly in their
interaction characteristics with \chem{H_2}). This cocrystal
formation enables the stabilization of surface-phase CO
molecules at temperatures far exceeding their typical triple
point, highlighting the critical importance of considering
multi-species interactions when modeling gas-phase depletion
processes in protoplanetary disks. \response{Experiments
  have also identified qualitatively similar trapping of
  adsorbates in similar situations \citep[e.g.][]
  {2016ApJ...825...89H}.
}

It should be further noted that while the primary
investigations in this work focus on single-layer adsorption
phenomena, additional studies of second-layer \chem{H_2O}
adsorption on silicate-supported water monolayers revealed a
single-molecule adsorption energy of $0.28~\eV$, resulting
from the combined effects of hydrogen bonding and van der
Waals interactions. When neighbor interactions are
incorporated, this corresponds to a ``multiple-layer
snowline'' at $\sim 1~\au$ as indicated by supplementary KMC
simulations (not directly illustrated in this work), a
finding consistent with conventional snowline
scenarios. 

Nevertheless, the single-layer adsorption conditions remain
particularly relevant for several reasons. First, grain
coagulation and subsequent planet formation processes are
predominantly governed by single-layer surface conditions,
as the sticking versus bouncing outcomes during collisions
are primarily determined by center-of-mass collision
energies that are most sensitive to the last adsorbed
molecular layer. Second, the fundamental physical mechanism
enabling CO condensation at elevated temperatures through
enhanced binding with \chem{H_2O} neighbors remains equally
valid in multiple-layer scenarios, as the crucial factor is
the strengthened binding energy and consequent confinement
of \chem{CO} molecules within the water ice matrix,
regardless of the total number of adsorbed layers.

\subsection{Molecule Condensation and Gas-Phase Depletion} 
\label{sec:co-condensation}

Our calculations indicate that the CO midplane snowline lies
significantly closer to the central star than conventional
estimates suggest. In a representative PPD around a
solar-type star with $T(1~\au)\simeq 150~\K$, the CO
snowline is expected at around $30~\au$
\citep[e.g.,][]{Oberg2023}. When adopting the higher
midplane temperature used in our disk model ($280~\K$
instead of $150~\K$ at $1~\au$), this discrepancy becomes
even more pronounced. This inward shift of CO snowline in
our model mainly arises from enhanced \chem{CO}--\chem{H_2O}
binding, which promotes earlier condensation of CO onto
water-rich grain surfaces.

Observationally, molecular snowline locations can be
inferred from chemical tracers of gas-phase abundances, such
as \chem{N_2H^+} for CO, which usually place the CO snowline
beyond 40\,AU in disks around solar-type stars
\citep{Qi2013,Qi2019}. However, these diagnostics are not
always uniquely linked to volatile depletion at the
corresponding snowline and can be affected by optical depth
effects, non-thermal desorption, or chemical processing
\citep{van'tHoff2017A&A...599A.101V}. These measurements
typically trace the traditional $20~\K$ condensation front,
where the CO gas abundance drops dramatically. In contrast,
our models predict the onset of CO condensation
significantly closer to the star, at a region not directly
captured by such observations. This suggests that a
substantial disk region may contain partially CO-ice-coated
grains, effectively extending the CO-ice zone inward
compared to classical estimates.
 
This expanded CO-ice zone has important implications for the
discrepancy between CO-based gas mass estimates and other
disk mass tracers. Early ALMA observations have often
inferred surprisingly low gas disk masses under the standard
assumption of a CO--\chem{H_2} ratio of $10^{-4}$
\citep[e.g.][]{Miotello2017, Long2017}. These estimates
often yield total gas masses below one Jupiter mass for
disks only a few Myr old, implying that giant planet
formation must terminate very early. One proposed solution
is that current thermochemical models underestimate CO
depletion, thereby underpredicting CO line fluxes. Indeed,
models incorporating grain-surface chemistry and CO
conversion into \chem{CO_2} ices can produce higher apparent
gas masses \citep{Ruaud2022}. However, such models still
fail to fully resolve the discrepancy; for example, the
CO-based gas mass estimate for IM Lup remains an order of
magnitude lower than its dynamical disk mass
\citep{Deng2025arXiv}. Our results should therefore be
incorporated in future models to improve the accuracy of
disk mass determinations, a key parameter for disk evolution
and planet formation models.

Our models also predict that \chem{H_2O} ice can sublimate
at lower temperatures and thus at larger disk radii, when it
is coated on carbonaceous grains. These shifts of snowline
locations have important implications for interpreting the
inner disk chemistry revealed by JWST MIR spectra. The
initial gas-phase C/O ratio in disks is largely determined
by the relative positions of major volatile snowlines. For
example, it is expected to be lower inside the \chem{H_2O}
snowline than in the region between \chem{H_2O} and
\chem{CO} snowlines \citep{Oberg11}. Current interpretations
of JWST spectra often invoke the sublimation of icy grains
as they drift inward and cross various snowlines
\citep[e.g.,][]{Mah2023}. Recent models that include CO
entrapment in water ice suggest elevated C/O and C/H ratio
inside the water snowline, by up to a factor of 10 over
$1~{\rm Myr}$ \citep{Williams2025}. This process is
analogous to the closer-in onset of CO condensation
predicted in our calculations. Precise knowledge of snowline
locations and their temperature dependencies is thus
essential for modeling disk chemistry. The outward shift of
the water snowline predicted here adds an additional layer
of complexity that future chemo-dynamical models should
consider.  Moreover, the lower sublimation temperature of
water ice implies that \chem{H_2O} vapor may persist at
colder temperatures, motivating future far-infrared
observations with facilities such as PRIMA to probe water
beyond the canonical snowline.

\section{Summary}
\label{sec:summary}

We have combined vdW-corrected \textit{ab-initio} DFT
calculations (\S\ref{sec:micro}) and kinetic Monte Carlo
simulations (\S\ref{sec:kmc}) to quantify how key volatiles
(\chem{H_2O}, \chem{CO}, and \chem{H_2}) adsorb onto
carbonaceous and silicate grain surfaces.  Although complete
representations of amorphous carbon and amorphous silicates
are computationally prohibitive, the surfaces of model
amorphous carbon and crystalline enstatite have been sampled
as proxies. The strong chemisorption energy
($\gtrsim 1~\eV$) of \chem{H_2O} and CO molecules on
silicate surfaces renders the prospective error from using
such proxies relatively less important. By sampling various
facets of crystalline enstatite, the bracketing approach
yields results for silicate surfaces that are reasonable,
semi-quantitative approximations to realistic amorphous
silicates.

The material-dependent adsorption energies, when
incorporated into the KMC framework, lead to distinct
surface coverage patterns and well-defined snowlines on
grain surfaces in a representative protoplanetary disk model
(Figure~\ref{fig:surface_coverage}). Our calculations have
also demonstrated a dichotomy between weak physisorption on
carbonaceous grains and strong chemisorption on silicate
grains (Tables~\ref{table:ads-nbr-carbon} and
\ref{table:ads-silicate}), which results in carbonaceous
grains losing their volatile coatings in the inner disk
while silicate grains remain molecularly coated over a much
wider radial range.

Building on this physical picture, we show that
history-dependent initial surface conditions and
evolutionary paths can shift snowline locations
(\S\ref{sec:kmc-init-cond}), that multi-species adsorption
and enhanced \chem{CO} binding in \chem{H_2O}-rich ices can
move the effective CO surface snowline inward and strengthen
gas-phase CO depletion (\S\ref{sec:kmc-cocrystal}), and that
these effects together impact disk gas-mass estimates, the
interpretation of snowline-sensitive tracers, and JWST
spectra in planet-forming disks
(\S\ref{sec:co-condensation}).

From an observational perspective, the carbonaceous snowline
on carbonaceous grain surfaces developed in this work should
influence carbon abundance patterns in extrasolar planetary
systems, as the lack of ``glue molecules'' on carbonaceous
grains at warm temperatures can lead to the generic scarcity
of carbon in regions closer to the host star.  Future
observational campaigns, particularly those utilizing
improved characterization of elemental abundances in
exoplanetary atmospheres, will provide crucial tests of
these predictions across a broader range of planetary system
architectures.  Such data may additionally offer constraints
on potential planetary migration histories. This work
establishes a foundation for investigating key astrophysical
processes through a modernized approach grounded in the
fundamental physics of molecules, atoms, and electrons,
bridging microscopic interactions with macroscopic
astrophysical phenomena.

\begin{acknowledgments}
  This work is supported by the NSFC General Project
  12573067. This research was enabled partially by the
  High-performance Computing Platform of Peking
  University. We thank our collegues: Di Li, Luis C. Ho,
  Jiao He, Jeremy Goodman, Ahmad Nemer, Satoshi Okuzumi,
  Kengo Tomida, Kazuyuki Omukai, Mordecai Mac-Low, Xiao Hu,
  Siyi Feng, for helpful discussions and comments.
\end{acknowledgments}

\appendix

\section{Thermodynamic Corrections for Adsorption
  Energies}
\label{sec:apdx-adsorb-therm-corr}

The thermodynamic treatment of adsorbate-substrate systems
requires careful consideration of their inherent complexity
compared to isolated molecular systems. Dust grain
substrates possess substantially larger heat capacities than
individual adsorbate molecules and maintain stiff thermal
equilibrium with the diffuse radiation fields characteristic
of interstellar media and protoplanetary disks
\citet{1997ApJ...490..368C}. Consequently, these substrates
effectively function as isothermal heat reservoirs during
adsorption-desorption processes. When molecular desorption
occurs, the thermal relaxation timescale in the gas phase
through collisions with other molecules can be estimated
following \citet{DraineBook}:
\begin{equation}
  \label{eq:tau-relax}
  \begin{split}
    \tau_{\mathrm{coll,gas}}
    & \equiv [n_{\chem{H_2}}\langle\sigma
      v_{\mathrm{th}}\rangle ]^{-1} \\ 
    & \sim 1~\s\times
      \left(\dfrac{T}{10^2~\K}\right)^{-1/2} 
      \left(\dfrac{\rho_{\mathrm{gas}}}
      {10^{-14}~\g~\cm^{-3}}\right)^{-1/2} .
  \end{split}
\end{equation}
Under typical protoplanetary disk mid-plane conditions,
$\tau_{\mathrm{coll, gas}}$ exceeds the desorption timescale
by more than ten orders of magnitude, effectively halting
thermal energy exchange between desorbed molecules and the
surrounding gas phase. This dramatic timescale separation
necessitates distinct thermodynamic treatments for different
molecular degrees of freedom: molecular motion perpendicular
to the substrate surface (the desorption direction) should
be treated as adiabatic, with the internal energy change
$\Delta\epsilon_{\ad}$ serving as the relevant
thermodynamic potential. Conversely, motion parallel to the
surface maintains continuous thermal contact with the
substrate heat reservoir, requiring that interactions with
neighboring adsorbates during hopping and desorption
processes be treated isothermally.

The vibrational characteristics of adsorbed molecules were
determined through analysis of their quasi-harmonic
oscillations near energy minima. Following convergence to
minimal energy configurations with fixed substrate ionic
structures, the VASP software package was employed to
evaluate second derivatives of the potential energy surface
via finite differences. Diagonalization of the resulting
Hessian matrices yielded eigenfrequencies for adsorbate
oscillation modes, enabling precise thermodynamic
corrections.

For the adiabatic dimension (perpendicular motion),
thermodynamic corrections primarily account for the
transition from free translational motion to confined
quasi-harmonic oscillations. The internal energy correction
incorporating zero-point energy (ZPE) contributions is given
by:
\begin{equation}
  \label{eq:de-perp} \delta
  \epsilon_\perp = h\nu \left( \dfrac{1}{2} +
    \dfrac{1}{e^{\beta h\nu} - 1} \right) -
  \dfrac{1}{2}\kb T\ ,
\end{equation}
where $\nu$ represents the vertical eigenfrequencies
tabulated in Tables~\ref{table:ads-nbr-carbon} and
\ref{table:ads-silicate}, and $\beta \equiv (\kb T)^{-1}$.

For the isothermal dimensions (parallel motion), the
Helmholtz free energy serves as the appropriate
thermodynamic potential. The binding energy correction is
derived from the free energy difference between the bounded
oscillatory state and the unbounded translational state:
\begin{equation}
  \label{eq:df-para}
  \begin{split}
    & f_{\mathrm{osci}} = \dfrac{h\xi}{2} + \dfrac{1}{\beta}\ln
                   \left[1 - e^{-\beta h \xi}\right]\ ,\\
    & f_{\mathrm{trans}} 
    = - \dfrac{2}{\beta} \ln \left[ \delta l \left(
        \dfrac{h^2\beta}{2\pi \mu} \right)^{-1/2} \right]\ ,
  \end{split}
\end{equation}
where $\xi$ denotes the horizontal eigenfrequencies from
Tables~\ref{table:ads-nbr-carbon} and
\ref{table:ads-silicate}, $\mu$ represents the molecular
mass, and
$\lambda_{\mathrm{th}}\equiv (h^2\beta/2\pi\mu)^{1/2}$
corresponds to the thermal de Broglie wavelength. The
parameter $\delta l$ describes the average intermolecular
distance on the substrate surface, which varies with both
gas-phase and adsorbed-phase conditions. For computational
simplicity, $\delta l\sim 10~\mathrm{\AA}$ was adopted,
approximately corresponding to the distance where
interactions between adjacent adsorbates become
negligible. The logarithmic dependence on $\delta l$ ensures
that the results remain insensitive to its exact value,
requiring only order-of-magnitude accuracy. Increasing
$\delta l$ elevates the value of $\delta f_\parallel$,
thereby generally weakening the binding between adsorbates.

\section{Construction of Dust Grain Substrate Models}
\label{sec:apdx-substrate}

\subsection{Graphene Substrate Construction}
\label{sec:apdx-carbon-graphene}

The potential energy surface (PES) for molecular
adsorption was systematically characterized through
computational mapping of adsorption energy curves for
\chem{H_2O}, \chem{CO}, and \chem{H_2} using a
single-layer graphene sheet as the model substrate. A
$3\times 3$ supercell configuration was employed,
comprising 18 carbon atoms arranged in a diamond-shaped
lattice with dimensions $7.40~\ang \times 7.40~\ang$ and
an open angle of $\pi/3$. To eliminate spurious
interactions between periodic images, a vacuum space of
$25~\ang$ was maintained above the graphene
surface. Adsorbate molecules were systematically
positioned at varying distances from the substrate surface
to sample the complete adsorption potential landscape.

Brillouin zone sampling was conducted using a
$\Gamma$-centered $1\times 1\times 1$ k-point grid during
initial DFT calculations. The adequacy of this sampling
scheme was verified through comparative tests employing
denser $3\times 3\times 1$ and $5\times 5\times 1$
$k$-point meshes at identified energy minima, confirming
numerical reliability and convergence of the computed
adsorption energies. Following initial energy curve
determination, full structural relaxation of adsorbate
configurations was performed to locate the true energy
minima and obtain accurate minimal adsorption energies.

\subsection{Amorphous Carbon Substrate Construction}
\label{sec:apdx-carbon-amorphous}

A representative model for interstellar amorphous carbon
grains was constructed through a multi-step computational
procedure. Initial configuration generation involved
placement of 75 carbon atoms on a randomly perturbed
$5\times 5\times 3$ lattice within a periodic boundary
condition unit cell. To emulate the hydrogenation conditions
expected in interstellar environments, 30 hydrogen atoms
were randomly distributed atop the carbon bulk,
preferentially associating with carbon atoms exhibiting low
coordination numbers.

Structural optimization was performed using DFT with full
consideration of electronic interactions, allowing both
ionic positions and supercell dimensions to vary during
relaxation. Convergence was achieved when the total force on
all atoms fell below the threshold of
$10^{-2}~\eV~\ang^{-1}$. Hydrogen atoms failing to form
chemical bonds with carbon during this process were
subsequently removed from the system, simulating the natural
saturation processes occurring in astrophysical
environments.

The final optimized structure possessed a supercell with
transverse dimensions of $(9.09~\ang)^2$ and exhibited a
mixed hybridization state distribution: 47\% of carbon atoms
in $sp^2$ configuration and 53\% in $sp^3$
configuration. The resulting carbon bulk demonstrated an
average thickness of $\sim 5.1~\ang$ and a density of
$\sim 2.4~\g~\cm^{-3}$, values consistent with
experimentally characterized amorphous carbon
materials. Surface hydrogen coverage reached
$\sim 0.16~\cm^{-2}$, corresponding to approximately 1\%
atomic hydrogen concentration for a spherical grain of
radius $0.05~\micron$, in good agreement with experimental
measurements \citet{1998A&A...332..291J}.

\subsection{Silicate Substrate Construction}
\label{sec:apdx-carbon-silicate}

Model silicate substrates were constructed using iron-poor
enstatite \chem{MgSiO_3} in the Pnma space group as a
representative magnesium silicate system. Surface selection
presented significant complexity due to the theoretically
infinite variety of possible crystal facet orientations. To
ensure computational tractability while maintaining physical
relevance, surfaces with the lowest Miller indices, (100),
(010), and (001), were selected as representative
substrates. Surface terminations were specifically chosen to
minimize dipole moments in the near-surface region,
enhancing numerical stability and physical realism.

The (001) surface orientation exhibited particular
complexity, manifesting two distinct termination types: the
$(001)^+$ surface exposing magnesium atoms and the $(001)^-$
surface exposing silicon atoms. This directional asymmetry
resulted from the non-centrosymmetric nature of the Pnma
crystal structure. In contrast, the (100) and (010) surfaces
did not exhibit such directional dependence due to inherent
symmetry properties.

Supercell dimensions were carefully optimized for each
surface orientation: $(9.96~\ang \times 13.96~\ang)$ for the
(100) surface, $(9.66~\ang \times 13.96~\ang)$ for the (010)
surface, and $(9.66~\ang \times 9.96~\ang)$ for both
$(001)^\pm$ surfaces. A consistent vacuum thickness of
$20~\ang$ was maintained above all surfaces to prevent
artificial interactions between periodic images and ensure
accurate simulation of isolated surface-adsorbate systems.

\response{
\section{Adsorption energy sampling and distribution over
  amorphous carbon model}
\label{sec:apdx-amorph-ene-rand}

Using the carbonaceous bulk constructed in
Appendix~\ref{sec:apdx-carbon-amorphous} as a representative
for the amorphous carbon substrate, the single-molecule
adsorption energy of \chem{H_2O} and \chem{CO} is sampled
over 50 ramdom transverse sites above the surface to emulate
the stochasticity of amorphous substrates. The interaction
energy curve at different vertical heights of the adsorbed
molecule, assuming that the molecule takes the configuration
minimizing the total energy, is calculated using vdW-DFT, on
which the lowest point yields the
$|\Delta \epsilon_{\rm ad}|$ on the site. The histograms of
$|\Delta \epsilon_{\rm ad}|$ distribution over sites are
presented in Figure~\ref{fig:amorph-ene-sample} for
\chem{H_2O} and \chem{CO}, respectively. It is obvious that
the expectation values of $|\Delta \epsilon_{\rm ad}|$ agree
with the data presented in Table~\ref{table:ads-nbr-carbon},
showing moderate widths (full-width half-maximum
$\sim 0.05~\eV$ for \chem{H_2O}, and $\sim 0.02~\eV$ for
\chem{CO}) that are consistent with the consequences in this
work.

\begin{figure}
  \centering
  \includegraphics[width=0.45\textwidth, keepaspectratio]
  {\figdir/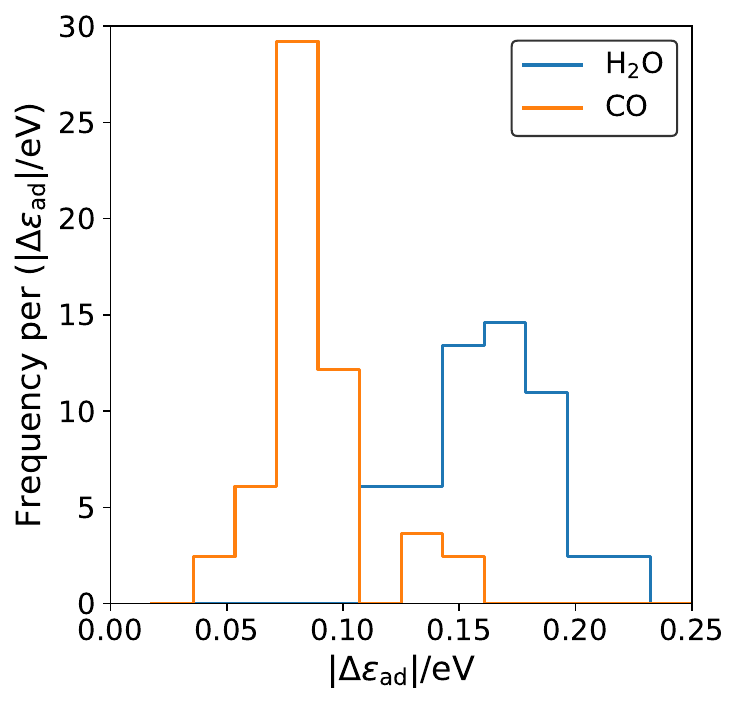}
  \caption{$|\Delta \epsilon_{\rm ad}|$ histograms for
    \chem{H_2O} and \chem{CO} for 50 random sites over the
    model amorphous carbon substrate (see
    Appendix~\ref{sec:apdx-amorph-ene-rand}). 
  }
  \label{fig:amorph-ene-sample}
\end{figure}

}

\section{General procedures of KMC simulations}
\label{sec:apdx-kmc-method}

KMC simulations have been performed extensively for various
studies for surface physical and chemical properties
\citep[e.g.][]{2019FrCh....7..202A}.  Three distinct
processes are modeled to occur on the simulated grain
surfaces, each characterized by specific rate equations:

Adsorption. Each unoccupied surface site experiences
molecular adsorption at a characteristic rate defined by
$\zeta_i \equiv n_i v_{{\rm th},i}\Sigma$, representing the
effective incoming flux frequency for molecular species
$i$. This formulation incorporates the gas-phase number
density $n_i$, the thermal velocity $v_{{\rm th}, i}$
derived from Maxwell-Boltzmann statistics, and $\Sigma$,
which denotes the effective adsorption cross-section with
intrinsic sticking probability implicitly included. Through
systematic analysis of solid-phase structures derived from
vdW-DFT calculations, an optimal value of
$\Sigma\simeq (3~\ang)^2$ is determined, providing a
consistent approximation applicable across all
substrate-adsorbate combinations examined in this study.

Hopping. Molecular migration across surface sites
is modeled through a hopping mechanism, where an adsorbate
occupying a site with $N$ adjacent vacancies exhibits a
hopping rate expressed as:
\begin{equation}
  \label{eq:freq-hop}
  k_{\rm hop} \simeq N \xi \exp \left[ \dfrac{\sum(\Delta
      \epsilon_{\rm hop} + f_{\rm osci}) - f_{\rm trans}}
    {\kb T} \right]\ ,
\end{equation}
The summation encompasses all neighboring sites, while
$f_{\rm osci}$ and $f_{\rm trans}$ represent the free energy
contributions from oscillatory and translational motions,
respectively, as defined in equation~\eqref{eq:df-para}. For
carbonaceous substrates, the hopping energy barrier
$\Delta \epsilon_{\rm hop} > 0$ induced by each neighbor is
quantified through the energy difference between adsorption
configurations with zero and one neighbors, as tabulated in
Table~\ref{table:ads-nbr-carbon}. In contrast, silicate
substrates exhibited significantly weaker inter-adsorbate
interactions relative to their substantial chemisorption
energies, permitting the approximation of the hopping
barrier simply by the absolute adsorption energy
$|\Delta \epsilon_{\ad}|$.

Desorption. Molecular detachment from surface sites
is characterized by a desorption rate governed by the
expression:
\begin{equation}
  \label{eq:freq-des}
  k_{\rm des}\simeq
  \nu \exp \left( \dfrac{\Delta \epsilon_{\ad} + \delta
      \epsilon_\perp - f_{\rm trans}} {\kb T} \right)\ ,
\end{equation}
where the vibrational frequency $\nu$ of adsorbed molecules
corresponds to the parameter defined in
equation~\eqref{eq:de-perp}. The desorption energy barrier
$\Delta \epsilon_{\ad}$ is computed using
adsorption energies from Tables~\ref{table:ads-nbr-carbon}
and \ref{table:ads-silicate}, with
appropriate adjustments accounting for both the quantity and
chemical identity of neighboring adsorbates to accurately
capture cooperative effects in molecular binding.

\bibliography{dust_cond}
\bibliographystyle{aasjournal}

\end{document}